\documentclass[aps,twocolumn,10pt,superscriptaddress,longbibliography]{revtex4-1} 
\usepackage{graphicx,subfigure}                           
\usepackage{physics}                                      
\usepackage{dsfont}                                       
\usepackage{amsfonts, amssymb, amsmath}		
\usepackage{charter}
\usepackage[colorlinks=true,linkcolor=blue,citecolor=red,urlcolor=magenta]{hyperref}
 
\usepackage{orcidlink}                                    

\begin{document}
	
\title{Discrete time quantum walk of locally interacting walkers}
\author{Vikash Mittal\;\orcidlink{https://orcid.org/0000-0002-4384-6992}}
\email{vikashmittal.iiser@gmail.com}
\affiliation{Department of Physics, National Tsing Hua University, Hsinchu 300044, Taiwan}
\author{Tomasz Sowi\'{n}ski\;\orcidlink{https://orcid.org/0000-0002-7970-4371}}
\email{tomasz.sowinski@ifpan.edu.pl} 
\affiliation{Institute of Physics, Polish Academy of Sciences, Aleja Lotnikow 32/46, PL-02668 Warsaw, Poland}    

\begin{abstract}
 In this work, we introduce a general form of a two-parameter family of local interactions between quantum walkers conditioned on the internal state of their coins. By choosing their particular case, we systematically study the impact of these interactions on the dynamics of two initially localized and noncorrelated walkers. Our general interaction framework, which reduces to several previously studied models as special cases, provides a versatile platform for engineering quantum correlations with applications in quantum simulation, state preparation, and sensing protocols. It also opens up the possibility of analyzing many-body interactions for larger numbers of walkers.
\end{abstract}
   
\maketitle

\section{Introduction}
The control of quantum correlations and spatial distributions in multi-particle systems represents a fundamental challenge in quantum physics, with direct implications for quantum technologies ranging from quantum control~\cite{Maciej_2023, Bryan_2023} to quantum computation~\cite{Nielsen_Chuang_2010}. Discrete-time quantum walks~\cite{Aharonov1993, Ambainis2001, Kempe2003, Renato2013, Kadian2021} have emerged as a powerful paradigm for achieving such control, leveraging quantum superposition and interference to generate non-classical transport phenomena~\cite{Ambainis2005, Chakraborty2020, Mittal2024} and several quantum resources~\cite{Abal2006, Carneiro2005, Alberti2015, Angelo2019, Mittal2025, Mittal2025a} with no counterparts in classical systems. Unlike their classical counterparts, quantum walks exhibit ballistic spreading and can generate long-range correlations through purely quantum mechanical effects~\cite{Ambainis2001, Ambainis2005}.     Furthermore, quantum walks have proven to be instrumental in exploring various physical phenomena, including topological phases.~\cite{Kitagawa2010, Asboth2012, Mittal2021}, Bloch oscillations~\cite{2010WitthautPRA, 2013Cedzich, 2015PhysRevA, 2018Perez, 2020ArnaultPRA, 2024Wojcik}, and quantum metrology~\cite{Matteo2022, Jex2023, Matteo2024, Chandru2024, Mostafa2025} to name a few.

When extended to multiple interacting particles, these systems become rich testbeds for exploring many-body quantum phenomena, including entanglement generation~\cite{VenegasAndraca2005}, spatial bunching~\cite{Angelo2019}, and localization transitions~\cite{Stefanak2011, Ahlbrecht2012}.

The study of interacting quantum walkers addresses a fundamental question on how controllable interactions between quantum walkers affect their spatial and entanglement dynamics. This question has both theoretical significance for understanding quantum correlation mechanisms and practical importance for developing quantum technologies that exploit controlled multi-particle interactions. Recent experimental advances in trapped-atoms systems~\cite{Jacob2024} and photonic~\cite{Sansoni2012, Silberhorn2014, Klauck2021, Esposito2022} quantum walks have made precise control over particle interactions achievable, motivating systematic theoretical investigations of controllable quantum correlations in discrete-time systems.

Previous theoretical work has established the foundation for interacting quantum walks, beginning with graph isomorphism applications~\cite{Berry2011} and extending to directional correlations~\cite{Stefanak2011}, bound state formation~\cite{Ahlbrecht2012}, and interaction-induced transitions between bosonic and fermionic behaviors~\cite{Wang2016}. These studies have been extended to include dynamic disorder~\cite{Toikka2020, Sergej2020}. However, a systematic study of how phase-controlled interactions affect both spatial distributions and quantum correlations remains incomplete. In addition to that, a unifying framework that encompasses previously studied interaction models as special cases is lacking. 

We address these gaps by introducing a general framework for local interactions between quantum walkers given by an arbitrary unitary transformation of the joint coin space when walkers coincide on the same lattice site. In our work, we focus mostly on a specific model of interaction, which is parameterized by a single phase parameter, rather than conducting a comprehensive exploration of the interaction given by a general unitary operator. This phase parameter modulates the interaction strength, allowing us to explore a wide range of behaviors, from non-interacting to strongly interacting regimes. Using this phase-tunable interaction, we investigate how the walkers' probability distributions evolve over time and how their spatial correlations can be engineered dynamically. 

Our general interaction framework reduces to several previously studied models as limiting cases. For instance, for a specific choice of parameters, we recover the position-only phase interactions~\cite{Ahlbrecht2012, Elias2018}, while for a different choice, we connect to the models studied in~\cite{Wang2016, Portugal2019}. This unifying perspective reveals that coin-state selectivity is a key organizing principle for understanding the dynamics of multi-walker correlations. It should be noted that similar interactions have been studied earlier~\cite{Stefanak2011, Ahlbrecht2012, Elias2018, Asaka2025}, although not in the same model or context.

Our work is organized as follows. In Section~\ref{sec:dtqw}, we remind the theoretical framework for discrete-time quantum walks and set the notation. In Sec.~\ref{sec:interaction}, we introduce a general interaction operator and we derive its special cases having natural physical interpretation. We then analyze marginal probabilities, lattice partitioning, and quantum correlations to quantify interaction-induced redistribution in Sec.~\ref{sec:quant}. Finally, in Sec.~\ref{sec:conclude}, we discuss implications, applications of our results, and conclude.

\section{Discrete-Time Quantum Walk}
\label{sec:dtqw}
A discrete-time quantum walk models the unitary evolution of a quantum particle (the walker) on a one-dimensional lattice. This evolution is controlled by the internal quantum states of the walker, which are usually taken to be two-dimensional (the coin) for one-dimensional lattices. Therefore, the total Hilbert space of a single walker is given as
\begin{equation} 
\label{HilbertSpace}
\mathcal{H} = \mathcal{H}_\mathtt{Coin} \otimes \mathcal{H}_\mathtt{Lattice},
\end{equation}
where $\mathcal{H}_\mathtt{Coin}$ is the coin space spanned by two internal degrees of freedom $\ket{\sigma}$, $ \sigma \in \{\ket{\uparrow},\ket{\downarrow}\}$, while $\mathcal{H}_\mathtt{Lattice}$ is the position space spanned by vectors $\ket{x}$, $x \in \mathds{Z}$. It means that at any moment $t$, the quantum state of the walker $\ket{\psi(t)}$ can be  written as the superposition
\begin{equation}
    \ket{\psi(t)} = \sum_{x, \sigma} \psi_\sigma(x; t) \ket{\sigma} \otimes \ket{x},
\end{equation}
where $\psi_\sigma(x; t)$ has a natural interpretation of probability amplitude of finding a walker at site $x$ in the internal state $\sigma$. The evolution proceeds through repeated applications of a unitary operator $U = S C$ composed of a coin flip $C$ followed by a conditional shift $S$. In this work, we consider the Hadamard coin, which in the Hilbert space ${\cal H}$ is represented by the operator  
\begin{equation}\label{equ:hadamard}
    C = \frac{1}{\sqrt{2}} 
\begin{pmatrix}
1 & 1 \\
1 & -1
\end{pmatrix} \otimes \mathds{1}.
\end{equation}
Clearly, this operator acts non-trivially only on the coin degrees of freedom. The conditional shift operator $S$ shifts the walker to the neighboring site of the lattice in the direction dependent on the state of the coin. Mathematically, it is written as
\begin{equation} \label{ShiftOperator}
    S = \sum_{x} \dyad{\uparrow} \otimes \dyad{x+1}{x}  + \dyad{\downarrow} \otimes \dyad{x-1}{x}.
\end{equation}
Consequently, this operation entangles the coin and position degrees of freedom during the evolution and introduces quantum interference into the walk dynamics.

For a given initial state of the walker $\ket{\psi(0)}$, the state after $t$ discrete time steps evolves as
\begin{equation}
    \ket{\psi(t)} = U^t \ket{\psi(0)}.
\end{equation}

Now we consider an extension of the problem to the case of two walkers moving on the same lattice. For further convenience, we write the joint Hilbert space of such a system again as Eq.~\eqref{HilbertSpace}. In this case, however, corresponding ingredients have a form
\begin{subequations}
\begin{align}
{\cal H}_\mathtt{Coin} &= {\cal H}_\mathtt{Coin}^{(1)}\otimes {\cal H}_\mathtt{Coin}^{(2)}, \\
{\cal H}_\mathtt{Lattice} &= {\cal H}_\mathtt{Lattice}^{(1)}\otimes {\cal H}_\mathtt{Lattice}^{(2)}.
\end{align}
\end{subequations}
With this notation, the evolution of non-interacting walkers is therefore governed by the joint operator
\begin{equation}
U_0 = U^{(1)} \otimes U^{(2)} = S^{(1)} C^{(1)} S^{(2)} C^{(2)}
\end{equation}
and at any time step $t$, the state of the composite system can be written as
\begin{equation} \label{EvolutionFree}
\ket{\Psi(t)} = U_0^t \ket{\Psi(0)}.
\end{equation}
Of course, again, the state $\ket{\Psi(t)}$ can be decomposed in the natural basis of two-walker Hilbert space ${\cal H}$ by introducing a two-walker wave function $\Psi_{\sigma_1,\sigma_2}(x_1,x_2; t)$ as
\begin{equation}
|\Psi(t)\rangle =\sum_{\sigma_1, \sigma_2}\sum_{x_1, x_2} \Psi_{\sigma_1,\sigma_2}(x_1,x_2; t) |\sigma_1,\sigma_2\rangle|x_1,x_2\rangle.
\end{equation}
\begin{figure}
    \centering
    \includegraphics[width=\linewidth]{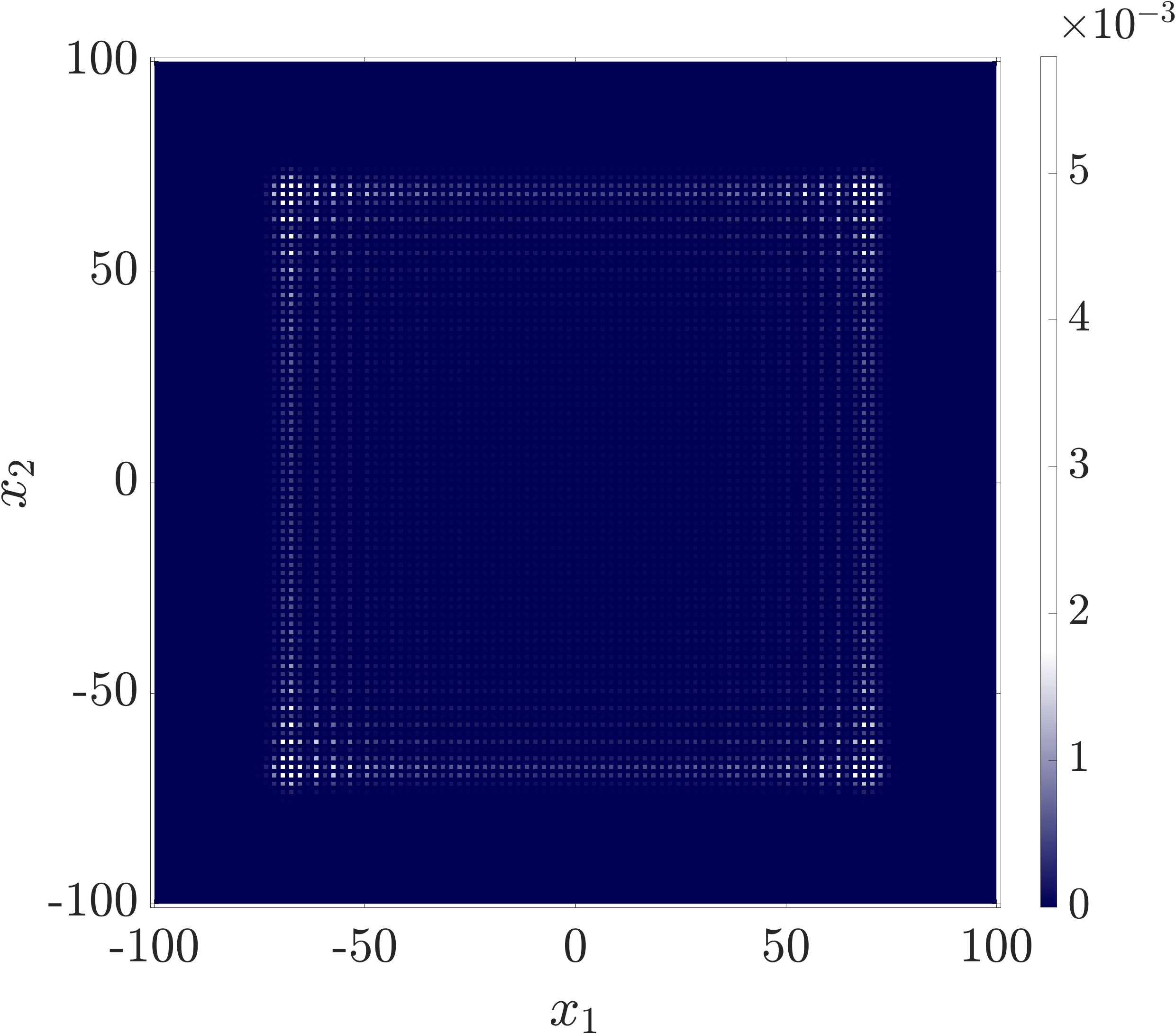}
    \caption{The two-particle density distribution $\mathrm{P}(x_1,x_2;t)$ after $t = 100$ time steps for non-interacting walkers ($\theta=0$). The initial state is taken to be $\ket{\Psi(0)} = \ket{+} \otimes \ket{+} \otimes \ket{0} \otimes \ket{0}$. }
    \label{Fig1}
\end{figure}
As an example, let us consider the time evolution of two walkers initially occupying the central site of the lattice and having the same and uncorrelated state of their coins. Namely, in the state
\begin{equation}
\label{eq:initialstate}
\ket{\Psi(0)} = \ket{+} \otimes \ket{+} \otimes \ket{0} \otimes \ket{0}.
\end{equation}
with $\ket{+} = (\ket{\uparrow} + i\ket{\downarrow})/\sqrt{2}$.
In Fig.~\ref{Fig1} we display the two-walker probability distribution 
\begin{equation} \label{JointProb}
\mathrm{P}(x_1,x_2;t) = \sum_{\sigma_1\sigma_2} \abs{\Psi_{\sigma_1,\sigma_2}(x_1,x_2;t)}^2
\end{equation}
of finding the walkers in the lattice independently of their spin after $t=100$ time steps. It is clear that the distribution exhibits distinctive peaks at the four corners, and the overall square-like symmetry is a hallmark of independent quantum walks. Quantum interference effects lead to higher probabilities at the extreme positions and strongly suppressed probabilities in the center and off-diagonal elements. In the subsequent sections, we introduce an interaction between the walkers, and this result will serve as a reference for those scenarios.

\section{Local interaction of two walkers}
\label{sec:interaction}
In order to analyze the dynamics of interacting particles, let us start by considering a reasonably general interaction operator that acts conditionally when both walkers occupy the same lattice site. Mathematically, this kind of operator can be written as 
\begin{equation}
    \label{eq:interaction}
    V = \mathds{I} + (\mathcal{O} - \mathds{1} \otimes \mathds{1}) \otimes \Pi
\end{equation}
where $\mathds{I}$ represent identity operator in the joint Hilbert space ${\cal H}$. The operator $\mathcal{O}$ is an arbitrary unitary transformation acting on the joint coin space ${\cal H}_\mathtt{Coin}$, while the operator $\Pi$ acts in the joined position space ${\cal H}_\mathtt{Lattice}$ and projects onto the subspace where both walkers occupy the same lattice site
\begin{equation}
    {\Pi} = \sum_{x_1,x_2} \delta_{x_1,x_2}\dyad{x_1,x_2}.
\end{equation}
By construction, the interaction operator $V$ is unitary and has a clear physical interpretation. The quantum state of the walkers is not changed when the walkers are at different positions. However, when both walkers occupy the same lattice site, the interaction operator $\mathcal{O}$ is applied to their coin states, keeping their positions unchanged. 

Without loss of generality, one can always express the unitary operator $\mathcal{O}$ with some hermitian operator $H$ as $\mathcal{O} = \mathrm{e}^{iH}$. Since it acts on the two-qubit Hilbert space, it can further be decomposed using Pauli matrices as
\begin{equation}
    \label{eq:hamiltonian}
    H = \sum_{i,j = 0}^3 h_{ij} \sigma_i \otimes \sigma_j
\end{equation}
where $\sigma_0$ is the identity $2\times 2$ matrix and $\sigma_i$ is the $i$-th Pauli matrix with real coefficients $h_{ij}$ which define the operator $H$ uniquely. 

While the general framework of Eq.~\eqref{eq:interaction} along with Eq.~\eqref{eq:hamiltonian} encompasses a broad class of local interactions, here we focus on a particularly important subclass of interactions that preserve the individual coin states of the walkers and introduce only an additional relative phase shift (it can be viewed as a counterpart of the interaction energy). This subclass is obtained by restricting $H$ to operators of the form
\begin{equation}
    H = \alpha \sigma_0\otimes\sigma_0 + \beta \sigma_3 \otimes \sigma_3, 
\end{equation}
where $\alpha$ and $\beta$ are two arbitrary real numbers. It is extremely  useful to parametrize parameters $\alpha$ and $\beta$ with two angles $\theta_\pm = \alpha \pm \beta$ since then the interaction operator $V$ can be written as 
\begin{equation} \label{VDef}
    V(\theta_+,\theta_-) = \mathds{I} + z_+ P_+ + z_-P_-
\end{equation}
where $z_\pm = \mathrm{e}^{i \theta_\pm} - 1$ and 
\begin{equation}
    \begin{aligned}
        P_+ &= \sum_{\sigma} \dyad{\sigma,\sigma} \otimes \Pi, \\
        P_- &= \sum_{\sigma_1\neq\sigma_2} \dyad{\sigma_1,\sigma_2} \otimes \Pi.
    \end{aligned}
\end{equation}
The projection operator $P_+$ ($P_-$) cuts from the joint Hilbert space of two walkers only these states where both particles occupy the same lattice site and have the same (opposite) coin states. Therefore, the interaction operator represented by operator $V$ can be viewed as an on-site spin-dependent interaction term preserving the internal state of individual walkers. 

This choice of interaction is motivated by both experimental feasibility and the desire to isolate the role of phase interactions in correlation dynamics~\cite{Sansoni2012, Schreiber2012, Manouchehri2014, Pan2019, Simon2020, Weiner2020, Qiang2024}. A systematic exploration of more general unitaries, including coin-flipping interactions, remains an interesting direction for future work.

Note that the collision-phase model studied in~\cite{Ahlbrecht2012} can be straightforwardly recovered by \eqref{VDef} by choosing $\theta_- = \theta_+$, {\it i.e.}, when both coin-state configurations acquire identical phases upon spatial coincidence. While in~\cite{Ahlbrecht2012} the authors focused on the spectral properties and molecular bound-state formation arising from collision phases, our work takes a complementary approach. We aim to systematically investigate how the interaction parameter controls spatial probability distributions and bipartite entanglement.

\section{Influence of interactions}
\begin{figure}
    \centering
    \subfigure{\includegraphics[width=0.48\textwidth]{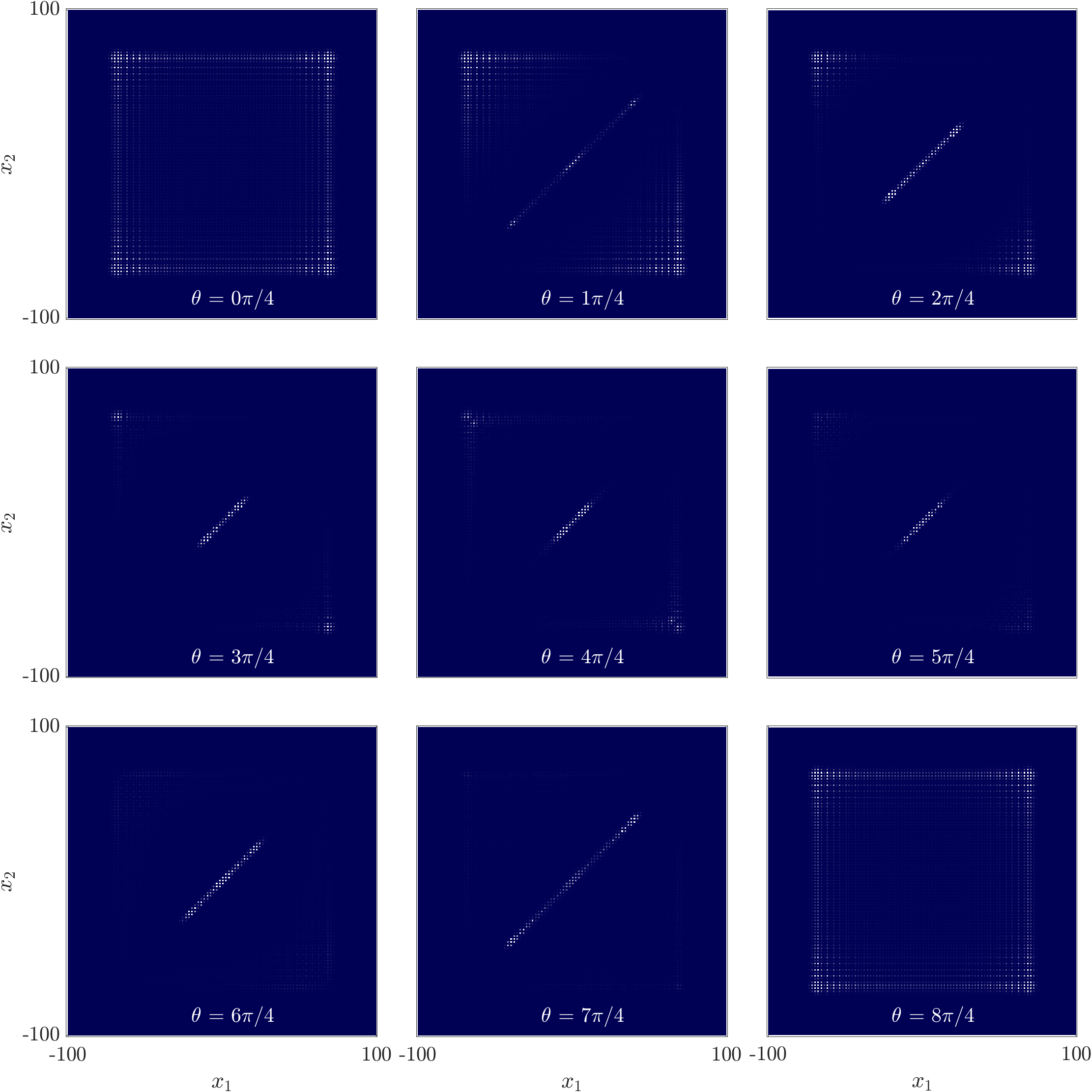}}
    \caption{The two-particle density distribution $\mathrm{P}(x_1,x_2;t)$ after $t = 100$ time steps for different interaction parameter $\theta$. By comparison with Fig.~\ref{Fig1} we investigate the impact of interactions to the dynamics. The limits for the colorbars are set to be $[0 \; \;0.006]$.}
    \label{Fig2}
\end{figure}
\label{sec:influence}
To show clearly how local interactions influence the dynamics of two walkers, let us focus on the simplest non-trivial case when walkers interact only when they have exactly the same internal state. It means that in Eq.~\eqref{VDef} we set $\theta_-=0 \implies z_- = 0$ and we consider the interaction operator of the form
\begin{equation}
    V_0(\theta) = \mathds{I} + (e^{i\theta} - 1) P_+,
\end{equation}
where, for simplicity, we omit the subscript of the angle $\theta$. In order to incorporate the effects of interactions into the dynamics, we modify the unitary evolution of the quantum walker in Eq.~\eqref{EvolutionFree} to the form
\begin{align}
\ket{\Psi(t)} &= (V_0(\theta)  U_0)^t \ket{\Psi(0)}.
\end{align}
It is clear that the value of the parameter $\theta$ influences the dynamics of the two walkers. Namely, for $\theta = 0$ one finds $V_0(0)=\mathds{I}$ and walkers evolve independently according to operator $U_0$. Along with increasing $\theta$, additional phase difference is introduced, and for $\theta = \pi$, this effect is maximal. Then, for $\theta>\pi$ interaction gradually returns to the non-interacting case. This effect is clearly reflected in the evolution of the joint probability Eq.~\eqref{JointProb}. In Fig.~\ref{Fig2}, we display the distribution $P(x_1,x_2)$ after $t=100$ steps for varying values of the interaction parameter $\theta$. We observe how the distribution changes between the square and the diagonal shape. As $\theta$ increases up to $\sim\pi/2$, we observe a subsequent accumulation of joint probability $\mathrm{P}(x_1,x_2)$ along the diagonal, showing that the largest is a probability of finding both walkers at the same position. Still, however, there is a non-vanishing probability of finding both walkers exactly on opposite sides of the lattice. Then, when angle $\theta$ is increased up to $\sim 3\pi/2$, the diagonal part of the correlation is amplified at the expense of the diminishing probability of finding walkers on opposite sides. These changes are gradual. Finally, for the largest angles, we see a rapid return to the square shape of the correlation corresponding to a non-interacting system. 

\begin{figure}
    \centering
    \includegraphics[width=\linewidth]{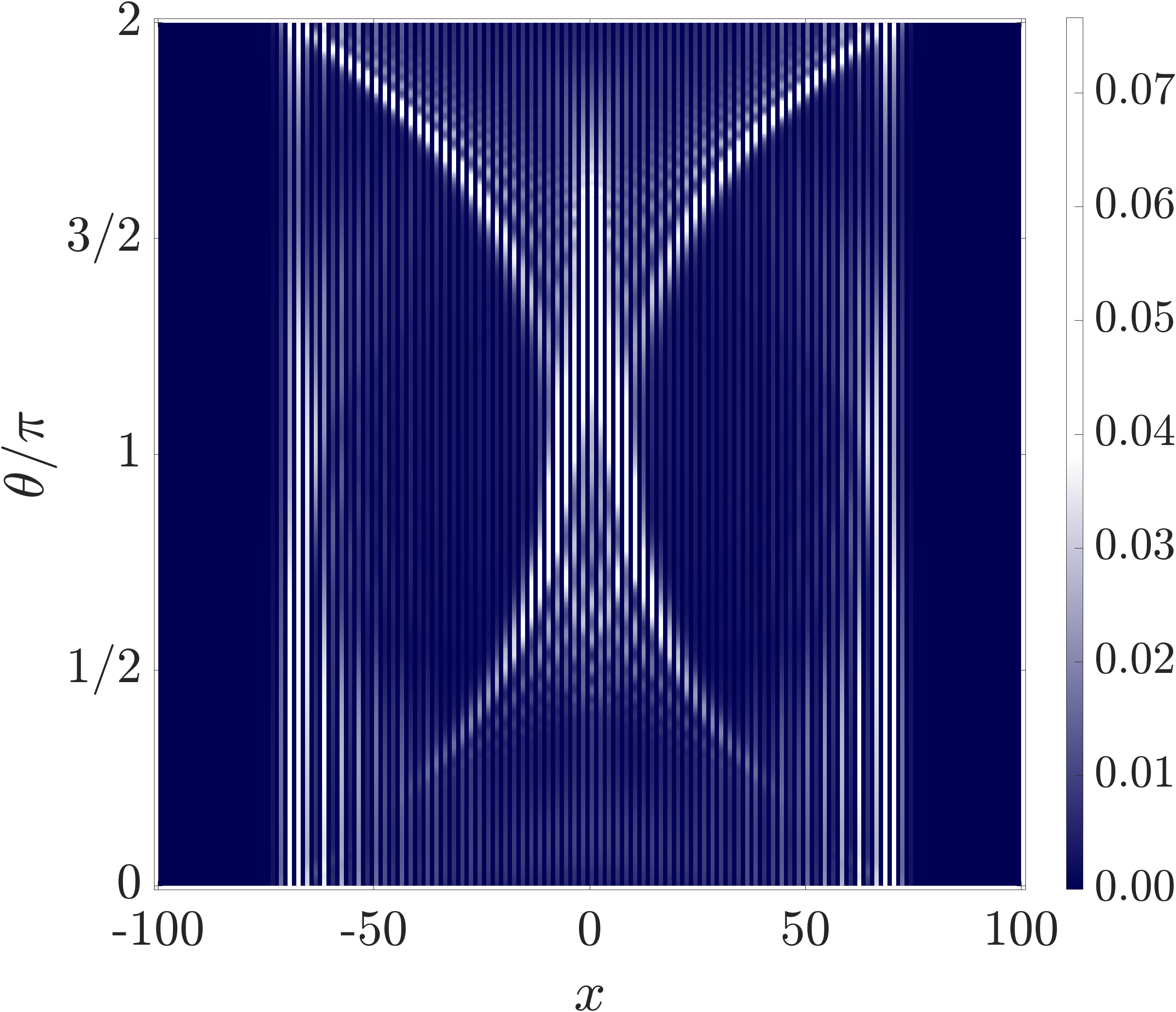}
    \caption{Density distribution of a single walker $n(x;t)$ after $t = 100$ time steps as a function interaction parameter $\theta$.}
    \label{Fig3}
\end{figure}
The changes in the two-body correlation described above are also well reflected in the probability distribution of a single walker $n(x;t)$ (the same for both walkers), defined as a marginal probability distribution
\begin{equation}
\label{eq:marginalP}
n(x;t) = \sum_{x'} \mathrm{P}(x,x';t)
\end{equation}
which we display in Fig.~\ref{Fig3}. Mentioned asymmetry in angle $\theta$ is clearly visible. We also observe a significant structural change around $\theta = \pi$ with the emergence of a central peak indicating a high probability of finding a walker in the center of the lattice. For larger $\theta$, the distribution of a walker becomes split and pushed out to the edges of the lattice. The rapid recovery at the edges is also evident from Fig.~\ref{Fig3}. 

It is worth noting the relationship between our observed spatial bunching and the molecular bound states discussed earlier in~\cite{Ahlbrecht2012}. In the previous analysis, bound states emerge as isolated eigenvalues in the spectrum of the evolution operator, resulting in exponentially localized wave functions in relative coordinates. Similarly, in the case studied here, the phase interaction induces strong spatial correlations, as evidenced by the pronounced diagonal features in the joint probability distribution (Fig.~\ref{Fig2}). However, a detailed spectral analysis lies outside the scope of the present work. Here, we focus on the time-dependent correlation dynamics and later entanglement generation, which provide complementary insights into interaction effects without requiring explicit construction of bound-state wave functions. Whether our interaction supports true bound states in the spectral sense, and how their properties depend on the coin-state selectivity, represents an interesting open question.

\section{Quantitative analysis}
\label{sec:quant}
In order to describe more quantitatively the dependence of correlations between walkers on the interaction angle $\theta$, we consider several complementary approaches. The first is related directly to the quantitative analysis of the two-walker density distribution, and it is based on a specific coarse-graining of the lattice. It is based on the observation that, due to the perfect initial localization of the walkers at site $x=0$ and because of the structure of the shift operator Eq.~\eqref{ShiftOperator}, it can be straightforwardly shown that for a fixed step $t$, the following relationship holds
\begin{equation}
\sum_{-t\leq x\leq t}\!\!n(x;t) = 1,
\end{equation}
{\it i.e.}, the position of each walker is limited to a segment of length $2t +1$. Therefore, it is natural to divide the lattice into three parts: the interior part ($\mathbf{I}$), containing all sites with $|x|\leq (t+1)/3$, and the left ($\mathbf{L}$) and right ($\mathbf{R}$) parts, which host the remaining sites with negative and positive indices, respectively. By calculating the total probability of finding a particle in each part of the lattice, we can estimate the probabilities of finding particles in the center or at the edges. Moreover, when applying this to the two-walker distribution density $\mathrm{P}(x_1,x_2)$, we can quantify the previously identified correlations by defining four probabilities that directly correspond to the likelihood of finding both walkers in the middle of the lattice (${\cal P}_\mathtt{A}$), on opposite edges (${\cal P}_\mathtt{B}$), on the same edge (${\cal P}_\mathtt{C}$), or with one walker in the middle and the other on an edge (${\cal P}_\mathtt{D}$). Exploiting the reflection symmetry ($\mathbf{L}\leftrightarrow\mathbf{R}$) of the evolution equation and the particles exchange symmetry of the distribution $\mathrm{P}(x_1,x_2)=\mathrm{P}(x_2,x_1)$ governed by the symmetry of the initial state, we can define them as
\begin{subequations} \label{PartProb}
\begin{align}
{\cal P}_{\mathtt{A}}(t) &= \sum_{x_1\in \mathbf{I}}\sum_{x_2\in \mathbf{I}} \mathrm{P}(x_1,x_2;t),\\
{\cal P}_{\mathtt{B}}(t)     &= 2 \sum_{x_1\in \mathbf{L}}\sum_{x_2\in \mathbf{R}} \mathrm{P}(x_1,x_2;t), \\
{\cal P}_{\mathtt{C}}(t) &= 2 \sum_{x_1\in \mathbf{L}}\sum_{x_2\in \mathbf{L}} \mathrm{P}(x_1,x_2;t), \\
{\cal P}_{\mathtt{D}}(t) &= 4 \sum_{x_1\in \mathbf{I}}\sum_{x_2\in \mathbf{L}} \mathrm{P}(x_1,x_2;t).
\end{align}
\end{subequations}
\begin{figure}
    \centering
    \subfigure{\includegraphics[width=\linewidth]{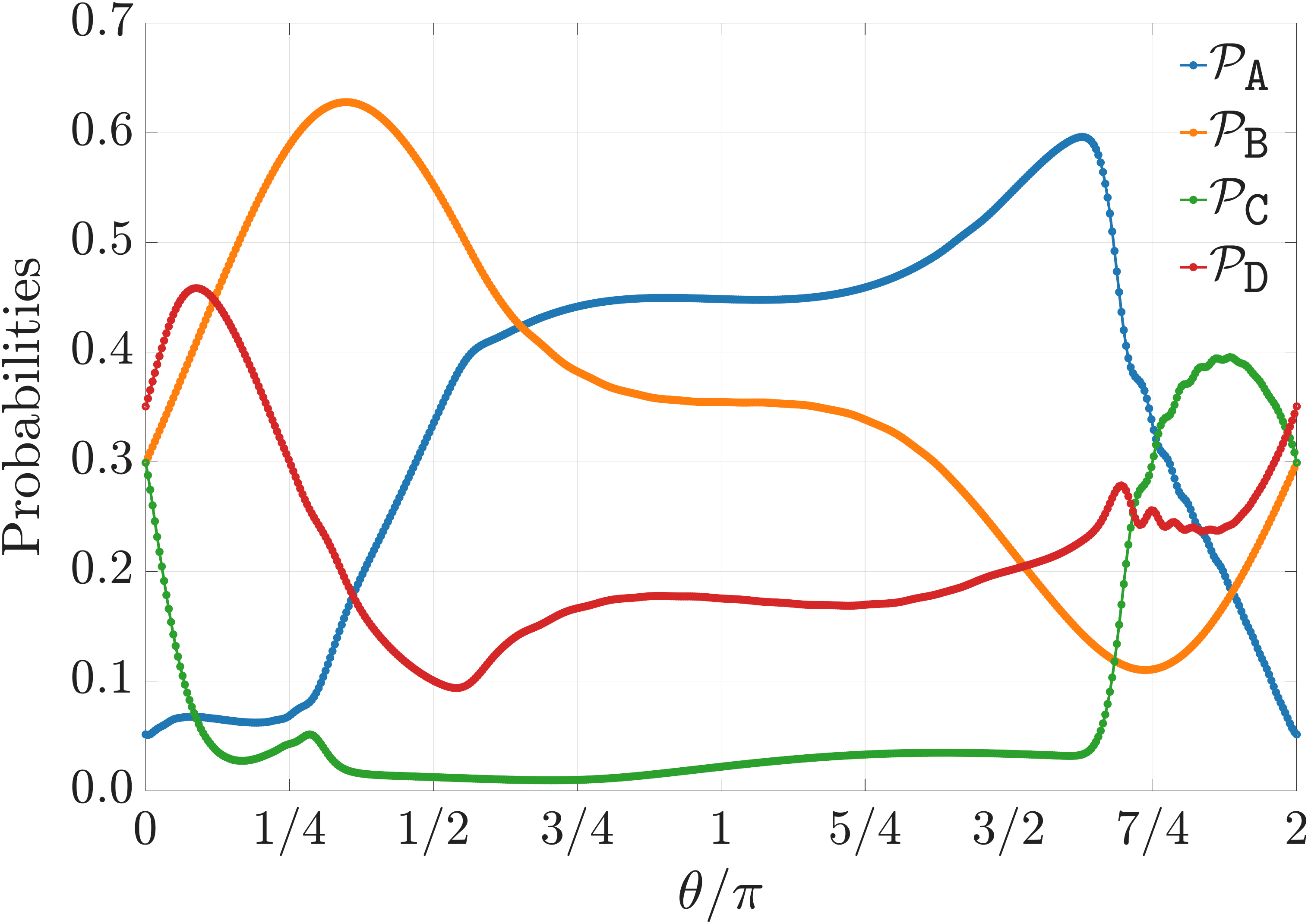}}
    \subfigure{\includegraphics[width=0.20\textwidth]{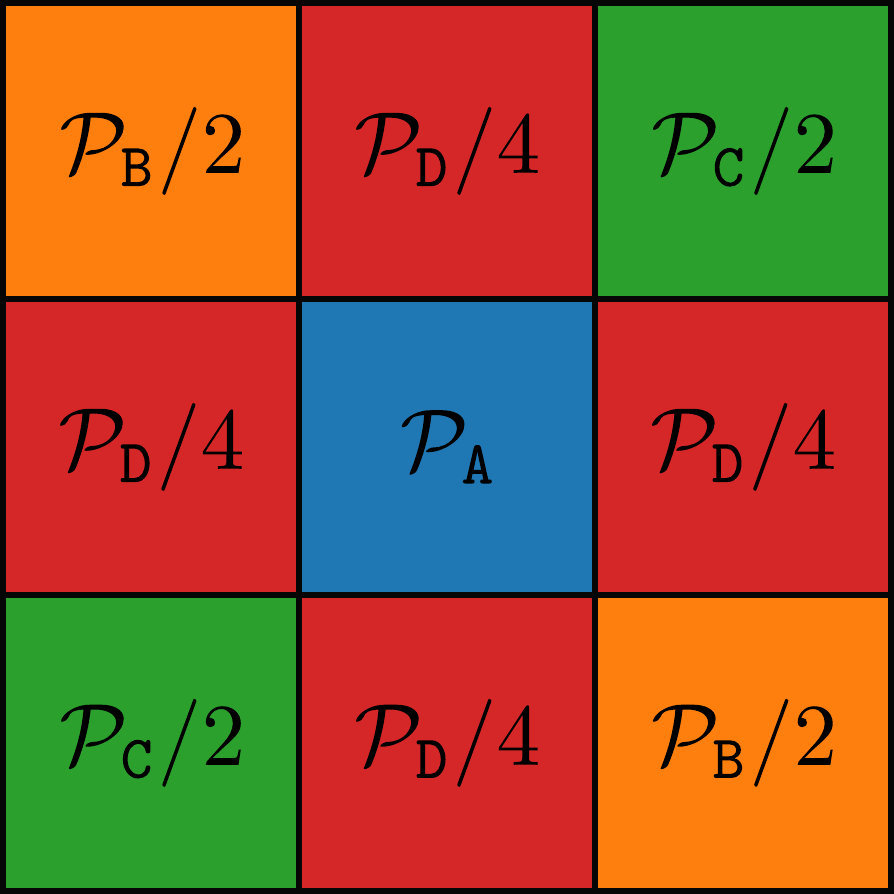}}
    \caption{Partial probabilities~\eqref{PartProb} of finding both walkers in different parts of the lattice obtained from the two-particle density distribution $\mathrm{P}(x_1,x_2;t)$ for $t=100$ as functions of the interaction parameter $\theta$. Different colors correspond to different areas of the lattice. For clarity, a schematic division of the probability distribution into different zones is shown below the plot.}
    \label{Fig4}
\end{figure}
In Fig.~\ref{Fig4} we present these four probabilities after $t = 100$ time steps as a function of interaction parameter $\theta$. For clarity, we also show a schematic representation of different probabilities after division of the lattice into three parts. First, we notice that for the non-interacting case ($\theta=0$) probability of finding both walkers on opposite sides of the lattice is equal to the probability of finding them on the same side ${\cal P}_\mathtt{B}={\cal P}_\mathtt{C}$. When we switch on the interactions, we observe the enhancement of opposite sides' configurations. Then, around $\theta \approx \pi/4$, a specific transition appears, which we also observed in the marginal probabilities -- probability quite rapidly transfers to the configurations with both particles occupying the central part of the lattice and saturates around $45\%$. This result is consistent with the appearance of a strong peak in the marginal probability presented in Fig.~\ref{Fig3}. When the interaction parameter $\theta$ crosses $3\pi/2$, all probabilities very quickly return to the values of the non-interacting scenario. For most of the range of the interaction parameter $\theta$, the probability of finding both the walkers on the same edge, ${\cal P}_\mathtt{C}$, remains quite low, and we observe a revival after $\theta = 3\pi/2$.

\begin{figure}
    \centering    
    \subfigure{\includegraphics[width=0.48\textwidth]{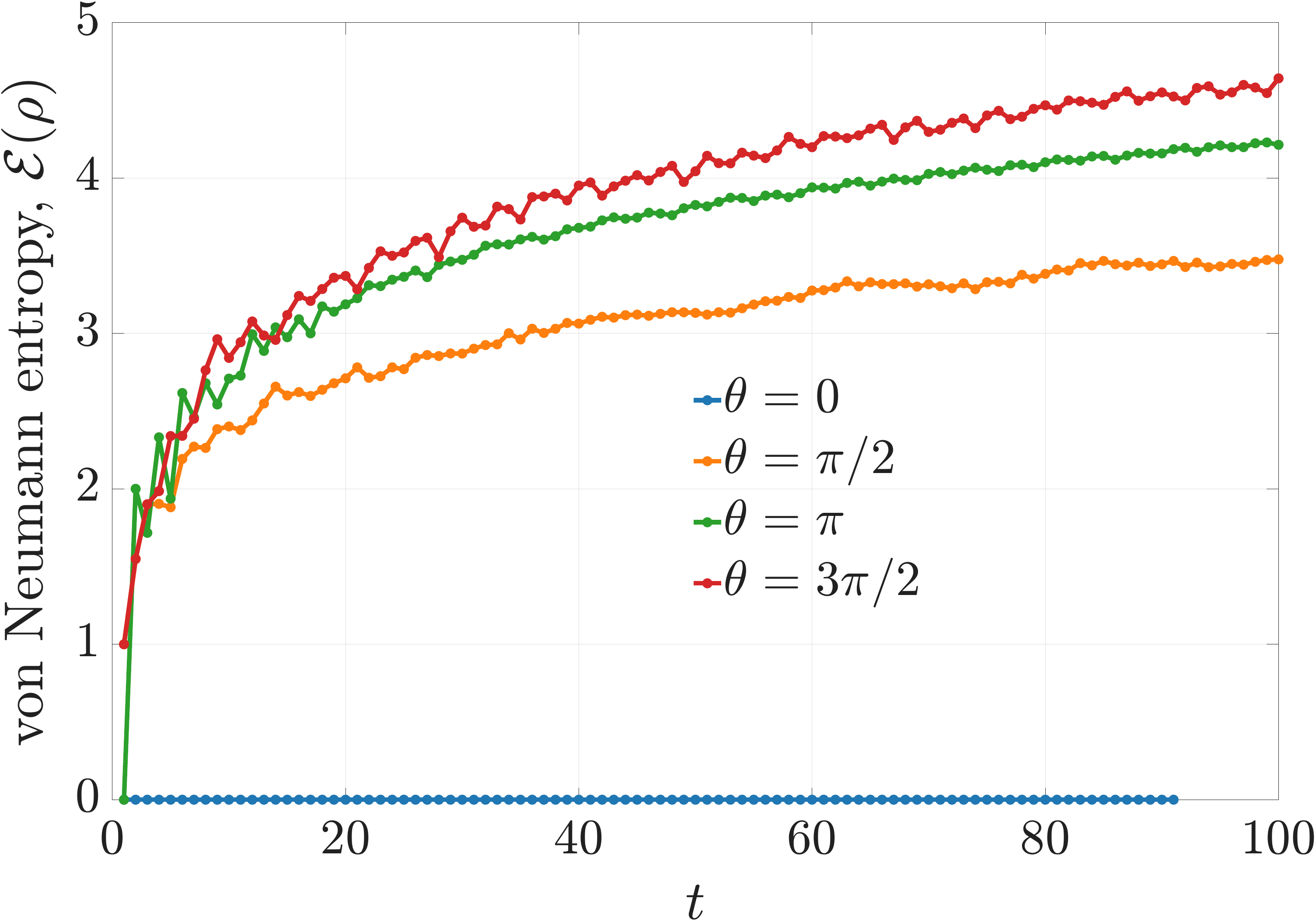}}
    \subfigure{\includegraphics[width=0.47\textwidth]{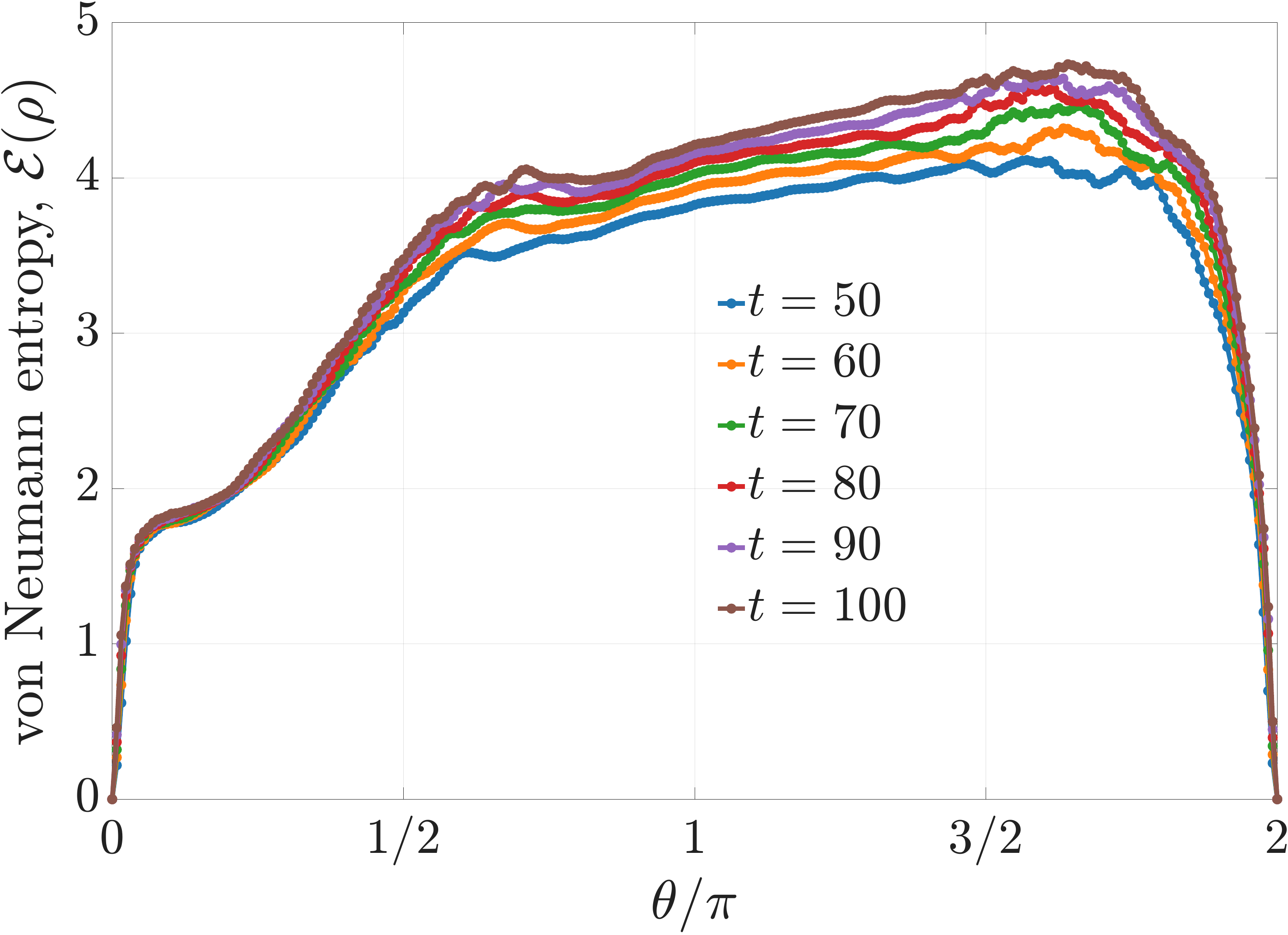}}
    \caption{Entanglement entropy $\mathcal{E}(t)$ between two quantum walkers as a function of time and interaction parameter $
    \theta$, respectively on (top) and (bottom) panel. The initial state of the composite system is given by Eq.~\eqref{eq:initialstate}.}
    \label{Fig5}
\end{figure}

The second, complementary approach of capturing mutual correlations between walkers is based on spectral decomposition of a single-walker reduced density matrix (due to the symmetry of the problem, it is the same for both walkers). The matrix is obtained by tracing out from the system's quantum state all degrees of freedom of the second walker. In the position-coin basis, it can be written as
\begin{multline}
\rho(x,\sigma;x',\sigma';t) \\=\sum_{\sigma_2}\sum_{x_2}\Psi^*_{\sigma,\sigma_2}(x,x_2;t)\Psi_{\sigma'\!,\sigma_2}(x',x_2;t).
\end{multline}
By performing spectral decomposition, one can express the matrix as
\begin{equation}
\rho(x,\sigma;x',\sigma';t) = \sum_i \lambda_i(t) \varphi_{i}^*(x,\sigma;t)\varphi_{i}(x',\sigma';t),
\end{equation}
where $\lambda_i(t)$ are its eigenvalues corresponding to an eigenorbital $\varphi_{i}(x,\sigma;t)$ obtained at given time step $t$. Of course, directly from the definition of the reduced density matrix, it follows that at any time step $t$, all the eigenvalues $\lambda_i(t)$ are non-negative reals and they sum up to unity, $\sum_i\lambda_i(t) = 1$. Therefore, one can easily quantify mutual correlations between walkers by calculating the von Neumann entropy defined as
\begin{equation}
{\cal E}(t) = -\sum_i \lambda_i(t)\log_2\lambda_i(t).
\end{equation}
Since initially the system is prepared in the initial state \eqref{eq:initialstate}, the reduced density matrix for $t=0$ has only one non-vanishing eigenvalue $\lambda_0=1$ corresponding to the orbital
\begin{equation}
\varphi_0(x,\sigma;t=0) = \frac{1}{\sqrt{2}}\left(\delta_{\sigma,\uparrow}+i\,\delta_{\sigma,\downarrow}\right)\delta_{x,0}.
\end{equation}
It means that initially the von Neumann entropy is equal $0$. We present the influence of interactions on the entropy in Fig.~\ref{Fig5}. The top panel presents the entanglement entropy between the two walkers as a function of time for fixed interaction strengths. For $\theta = 0$, we observe that entanglement between the two walkers is not produced. It is a straightforward consequence of the complete independence of walkers when interactions are switched off. For non-vanishing interactions, we observe small oscillations of the entropy at initial moments, which are followed by a saturation at long times. It is clear that the amount of entanglement produced significantly depends on the interaction parameter. This dependence is nicely captured in the bottom subplot of Fig.~\ref{Fig5}, which shows the entanglement entropy in the system as a function of $\theta$ for fixed time steps $t$. The entanglement exhibits clear $2\pi$-periodic behavior, directly reflecting the phase-based interaction mechanism. We observe the building up of the correlations in the system as we turn on the interaction parameter, followed by a plateau till some critical threshold around $\theta\approx 3\pi/2$. This again complements the pattern in the probability distribution. Once we cross this threshold, we observe a rapid decrease from this point onward for all the time scales. 

\begin{figure*}
    \centering
    \includegraphics[width=\textwidth]{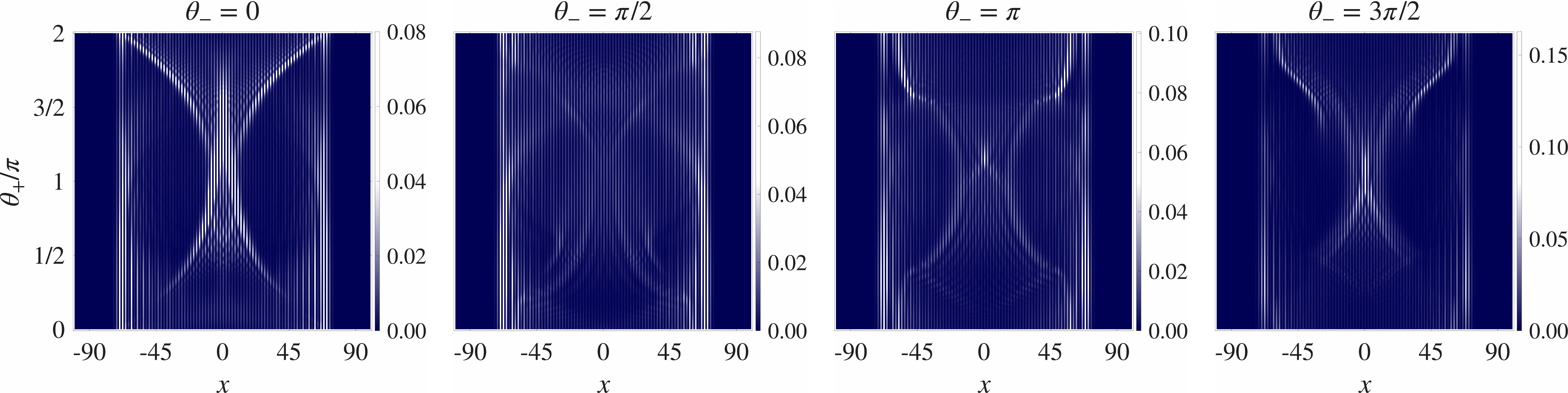}
        \caption{Density distribution of a single walker $n(x;t)$ after $t = 100$ time steps as a function of interaction parameter $\theta_+$ for different values of $\theta_- = 0, \pi/2, \pi, 3\pi/2$. Note that the case $\theta_- = 0$ corresponds to the single-parameter model visualized in Fig.~\ref{Fig3}.}
    \label{Fig6}
\end{figure*}

\section{Influence of opposite coin-state interaction channel}
\label{sec:extended}
To make the discussion more comprehensive, let us also briefly discuss the influence of the second channel of interactions controlled by the projection operator $P_-$ in \eqref{VDef}. In this way, we demonstrate the broader applicability of our framework and also capture the transition to the collision-phase model~\cite{Ahlbrecht2012}. For simplicity, we focus only on the density distribution of a single walker $n(x,t)$ after $t=100$ time steps. However, studying other, more sophisticated features of the system can also be easily performed. To illustrate the richer dynamics accessible with two independent parameters $\theta_\pm$, we fix $\theta_-$ at representative values $(0, \pi/2, \pi, 3\pi/2)$ and vary $\theta_+$ from $0$ to $2\pi$. As a result, we obtain counterparts of Fig.~\ref{Fig3} for different $\theta_-$, which we collect in Fig.~\ref{Fig6}.

We observe several notable features emerge from this extended analysis. The most prominent is that the density distribution $n(x,t)$ exhibits qualitatively different structures compared to the $\theta_- = 0$ case studied in previous sections. The interplay between same-spin and opposite-spin collision phases creates interference patterns that cannot be captured by any single-parameter models. Clearly, the distributions $n(x,t)$ exhibit complex, non-monotonic dependence on both parameters, reflecting the interference between different scattering channels in the joint coin space. These results demonstrate that our general framework provides access to a rich landscape of correlation phenomena beyond the single-parameter family analyzed in detail in the main text.

\section{Conclusion}
\label{sec:conclude}
In this article, we have presented a comprehensive study of discrete-time quantum walks of two interacting particles. We have introduced a quite general model of local interactions between walkers, and we demonstrated how a simple phase interaction can provide nontrivial dynamics of spatial probability distributions and quantum entanglement in the system.

Our results show that for the non-interacting case, the two walkers behave independently, spreading ballistically and localizing primarily near the edges of the lattice. As the interactions are introduced in the system, probability gradually shifts toward the central region of the lattice, indicating a collapse of the walkers' spread and the emergence of strong spatial correlations. At some specific value of the interaction parameter, we observe a transition in system behavior during which the probability distribution becomes heavily localized in the center, reflecting a high degree of bunching and interaction-driven localization. As we increase the strength of the interaction, the walkers tend to remain together. To quantify the effect of interactions further, we partition the lattice into several blocks at a fixed time step, and we perform coarse-grained calculations. This allows us to observe that the spatial distributions of interacting walkers shift from edge regions to the center of the lattice when the interaction parameter is varied. It confirms the strong influence of interactions on walkers' dynamics.

Next, we also investigate mutual correlations forced by interactions, quantifying them in terms of the entanglement entropy of the reduced density matrix of a walker. We observe a significant dependence of entanglement on the interaction parameter. The entanglement between the two walkers attains a maximum value at a particular interaction strength, demonstrating the system's ability to dynamically generate and control quantum correlations. Finally, we showed exemplary results for the general two-parameter model interpolating between the one-parameter model and the collision-phase model~\cite{Ahlbrecht2012}.

In the present study, we have limited ourselves to a very simple yet profound interaction. One of the future directions would be to consider more involves interactions and see the impact of them on the dynamics and the correlations in the system. The form of the most general interaction, which we discussed in the present article, can be used to gain physical insight into the system at the level of the Hamiltonian. 

Our results establish phase-controlled quantum walks as quantitative platforms for engineering spatial and entanglement correlations, with direct applications in quantum simulation of condensed matter systems, enhanced quantum sensing via controlled bunching, and scalable entanglement generation for quantum information processing.

\section*{Data availability statement}
All numerical data presented in this paper are available online~\cite{zenodo}.

\section*{Acknowledgment}
This research was supported by the National Science Centre (NCN, Poland) within the OPUS project No. 2023/49/B/ST2/03744 (TS). VM acknowledges a financial support by National Science and Technology Council (114-2811-M-008-075-MY2), MOST Young Scholar Fellowship (Grants No. 112-2636-M-007-008- No. 113-2636-M-007-002- and No. 114-2636-M-007 -001 -), National Center for Theoretical Sciences (Grants No. 113-2124-M-002-003-) from the Ministry of Science and Technology (MOST), Taiwan, and the Yushan Young Scholar Program (as Administrative Support Grant Fellow) from the Ministry of Education, Taiwan. For the purpose of Open Access, the authors have applied a CC-BY public copyright licence to any Author Accepted Manuscript version arising from this submission.

\end{document}